\documentclass[twocolumn,showpacs,preprintnumbers,amsmath,amssymb]{revtex4}

\usepackage{graphicx}
\usepackage{dcolumn}
\usepackage{bm}

\begin{document}


\title{Electronic band structure and Fermi surface of Ag$_5$Pb$_2$O$_6$}
\author{Tamio Oguchi}
\affiliation{%
Department of Quantum Matter, ADSM, Hiroshima University, Higashihiroshima 739-8530, Japan
}%

\begin{abstract}
We present electronic band structure of Ag$_5$Pb$_2$O$_6$ with layered hexagonal structure 
containing one-dimensional chains and two-dimensional Kagom\'{e} layers of silver. 
A half-filled conduction band shows extremely simple, single nearly-free-electron-like Fermi surface. 
The conduction band is composed of an antibonding state of Pb-$6s$ and O-$2p$ mixing with Ag-$4d$ and $5s$.  
Mass enhancement in the state density at the Fermi energy is expected to be negligibly small by comparing with 
the specific-heat data. 
Calculated Fermi velocity is consistent with small anisotropy observed in transport properties. 
Doping effects on the electronic structure are also discussed. 
\end{abstract}

\pacs{71.18.+y,71.20.Ps,72.15.Jf,74.25.Jb,74.70.Dd}

\maketitle

\section{Introduction}
Novel transport properties of oxide compounds have attracted much attention since the discovery of
the high critical temperature ($T_c$) superconductivity. 
Some perovskite manganites show colossal magnetoresistance~\cite{Tokura2000} and layered-structure cobaltates 
reveal large thermoelectric power.~\cite{Terasaki2002} 
It is widely believed that one-dimensional (1D) chains and/or two-dimensional (2D) 
layers consisting transition-metal and oxygen ions play a crucial role on the electronic properties in most of the cases. 
Recently Yonezawa and Maeno have observed anomalous $T^2$ dependence of the resistivity for 
single crystals of Ag$_5$Pb$_2$O$_6$.~\cite{Yonezawa2004} 
Ag$_5$Pb$_2$O$_6$ has a hexagonal crystal structure with 1D chains and  
2D Kagom\'{e} layers of silver ions interleaved with honeycomb layers of PbO$_6$ octahedra. 
The $T^2$ behavior of the resistivity is seen along both directions parallel and perpendicular to the 
layers in a wide temperature range. Anisotropy is found to be rather small like a factor of two in the resistivity. 
Measured specific-heat $\gamma$ value is just moderate, implying a small mass enhancement factor.
Yonezawa and Maeno have concluded that the $T^2$ dependence cannot be interpreted either 
by the electron-electron interaction or electron-phonon mechanism. 
In addition, they have reported an indication of a superconducting phase below 48mK, 
which has been confirmed by a resistivity measurement quite recently.~\cite{Yonezawa2005}  
Concerning the electronic structure of Ag$_5$Pb$_2$O$_6$, Brennan and Burdett have calculated 
the density of states and Fermi surface within a tight-binding (TB) model.~\cite{Brennan1994} 
It has been claimed that a carrier electron around the Fermi energy is delocalized over the entire 
silver 1D and 2D substructure. 
Recently, another band structure calculation has been reported with full-potential 
linearized muffin-tin orbital method.~\cite{Shein2005} Obtained band structure of Ag$_5$Pb$_2$O$_6$ 
is qualitatively different from the previous TB one and quite consistent with a result of the present study, 
as will be shown below. 

In this paper, we present electronic band structure and Fermi surface of Ag$_5$Pb$_2$O$_6$ 
calculated by using a first-principles density-functional method. 
Calculated results show a basic electronic structure of (Ag$^{+}$)$_5$(Pb$^{4+}$)$_2$(O$^{-2}$)$_6$+($-e$), 
namely a single nearly-free-electron-like conduction band with one electron per formula unit 
appearing in energy gaps formed in an ionic crystal. Very interestingly, the conduction band 
is found to be a 2D antibonding state of Pb-6$s$ and O-2$p$ and its dispersion along the $c$ direction originates in 
hybridization with Ag-$4d$ and $5s$, resulting in three-dimensional (3D) nearly-free-electron-like Fermi surface. 
In order to investigate transport properties, 
the Fermi velocity and Hall coefficients are evaluated from the obtained electronic band structure. 
Finally, doping effects for the Pb site on the electronic band structure are also discussed in detail. 

\section{Crystal Structure}
\begin{figure}[tbp]
\begin{center}
\includegraphics{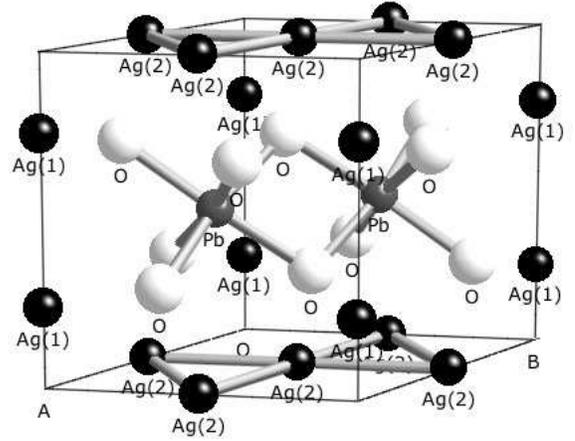}
\caption{\label{fig1} Crystal structure of the hexagonal Ag$_5$Pb$_2$O$_6$. 
Dark, grey and white spheres denote Ag, Pb and O atomic sites, respectively. 
}
\end{center}
\end{figure}

Crystal structure data of the hexagonal Ag$_5$Pb$_2$O$_6$ are taken from experiment by Jansen \textit{et al},~\cite{Jansen1990} 
as shown in  Fig.~\ref{fig1}. 
The space group is $P\bar{3}1m$ with lattice constants of $a=5.9324${\AA} and $c=6.4105${\AA}. 
Crystallographically independent atomic positions are given at (0,0,$z$) with $z$=0.2413 for 1D-chain silver (denoted as Ag(1)), 
(1/2,0,0) for 2D-layer silver (denoted as Ag(2)), (2/3,1/3,1/2) for Pb and ($x$,0,$z$) with $x$=0.6222 and $z$=0.6889 for O. 
There are two sites for Ag(1), three for Ag(2), two for Pb, and six for O in a hexagonal unit cell. 
Typical interatomic distances are 3.094{\AA} for Ag(1)-Ag(1), 2.966{\AA} for Ag(2)-Ag(2), 2.219{\AA} for Pb-O. 
It is quite important to note that the nearest atomic distances of Ag(1)-O and Ag(2)-O are 2.285{\AA} and 2.122{\AA}, 
respectively, which are both much closer than those of Ag(1)-Ag(1) and Ag(2)-Ag(2). 
Therefore, it is doubtful and unreliable to understand the electronic structure of 
Ag$_5$Pb$_2$O$_6$ only in terms of 1D chains and 2D layers of silver ions. 

\section{Methods}

The present first-principles calculations are based on the density-functional theory by
adopting the all-electron full-potential linear-augmented-plane-wave (FLAPW) method.~\cite{Weinert1981,Wimmer1981,Soler1989}  
Our implementation of the all-electron FLAPW method has been used successfully 
for a variety of condensed matter systems.~\cite{Oguchi1995,Iwashita1996,Shishidou2000,Oguchi2004}
Self-consistent-field (SCF) calculations are performed with the scalar-relativistic scheme and the 
improved tetrahedron integration method~\cite{Blochl1994} 
up to 16$\times$16$\times$16 k-mesh points in the Brillouin zone (BZ). 
Muffin-tin sphere radii are assumed to be 1.1{\AA} for Ag, 1.0{\AA} for Pb and 0.8{\AA} for O. 
Note that each partial density of states (DOS) shown below is projected on the corresponding partial spherical wave within the muffin-tin sphere. 
Exchange and correlation are treated within the local density approximation (LDA)~\cite{Janak1975} 
or generalized gradient approximation (GGA).~\cite{Perdew1996}  
For the present Ag$_5$Pb$_2$O$_6$ with the observed crystal structure, 
LDA and GGA give almost identical electronic band structure around the Fermi energy. We show only LDA results.  

\section{Results and Discussion}

\subsection{Electronic band structure}

\begin{figure}[tbp]
\begin{center}
\includegraphics{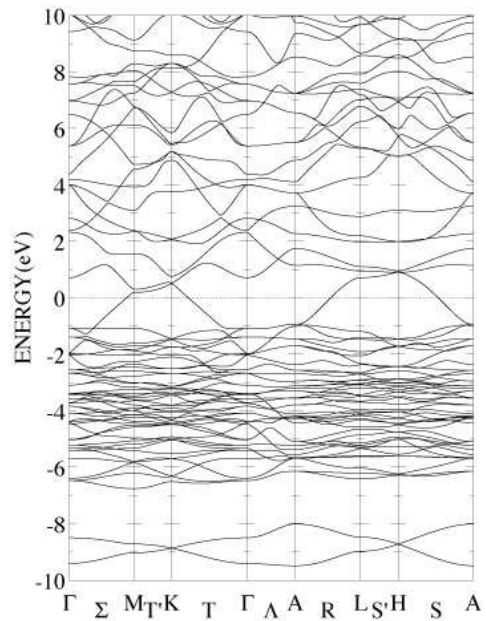}
\caption{\label{fig2} Calculated energy band structure of the hexagonal Ag$_5$Pb$_2$O$_6$. 
The Fermi energy is taken at the origin.}
\end{center}
\end{figure}

\begin{figure}[tbp]
\begin{center}
\includegraphics{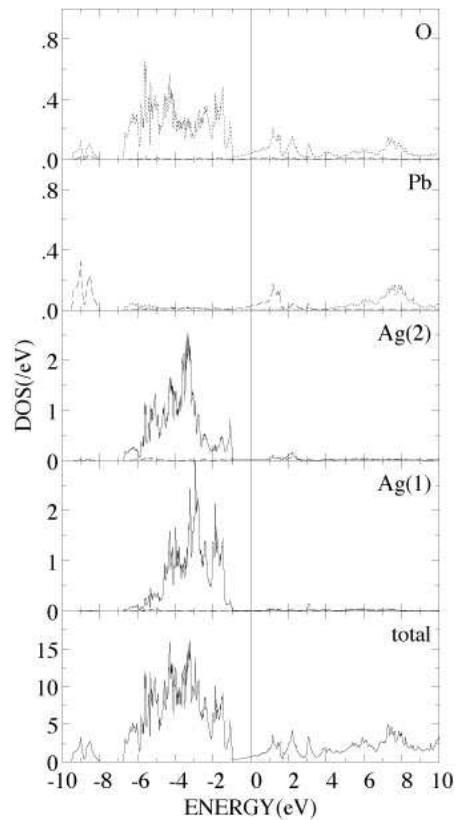}
\caption{\label{fig3} Calculated total and partial density of states (DOS) of the hexagonal Ag$_5$Pb$_2$O$_6$. 
The Fermi energy is taken at the origin. 
Broken, dotted and solid lines represent $s$, $p$ and $d$ angular-momentum components, respectively, in the partial DOS. 
}
\end{center}
\end{figure}

Figure \ref{fig2} shows calculated electronic band structure of the hexagonal Ag$_5$Pb$_2$O$_6$ along 
several high-symmetry lines in BZ. 
Each band dispersion is drawn according to its irreducible representation of the group of \textbf{k}. 
Two bands around $-9$eV in Fig.~\ref{fig2} are mostly bonding states between Pb-$6s$ and O-$2p$ orbitals. 
Complex bands between $-7$ and $-1$eV are composed of O-$2p$ and Ag-$4d$ orbitals mixed lightly with Pb-$6p$ and 
Ag-$5s$ as bonding states. 
A conduction band crossing the Fermi energy is dominated by an antibonding state of Pb-$6s$ and O-$2p$, 
forming a single cylindrical Fermi surface warping along the $c$ direction, as shown below. 
Those orbital components of each band can be seen more clearly in angular-momentum projected DOS 
shown in Fig.~\ref{fig3}. 
In the upper part of the antibonding bands around $+2$eV in Fig.~\ref{fig3}, one can see a peak containing Ag(2)-$4d$ and $5s$, 
which is a similar structure to that found in the previously calculated DOS.~\cite{Brennan1994} 
An important difference from the previous one is the existence of the antibonding band of Pb-$6s$ and O-$2p$ crossing the 
Fermi energy, which must govern the transport properties of the system. 
The Ag-$4d$ and $5s$ orbitals also hybridize with the antibonding band to some extent and the electron density at the 
Fermi energy is delocalized over the entire crystal. 
It is, therefore, summarized that the basic electronic band structure can be considered as 
(Ag$^{+}$)$_5$(Pb$^{4+}$)$_2$(O$^{-2}$)$_6$+($-e$), where
a conduction band forming a quasi-2D cylindrical Fermi surface with large warping is half occupied by one electron carrier. 

\begin{figure}[tbp]
\begin{center}
\includegraphics{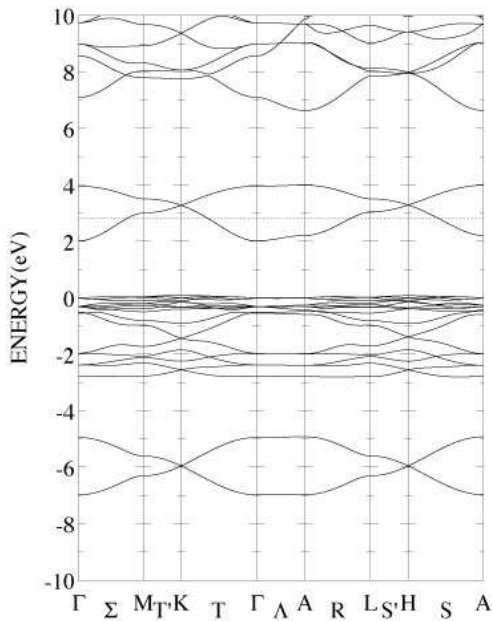}
\caption{\label{fig4} Calculated energy band structure of $\Box_5$Pb$_2$O$_6$ ($\Box$ = Ag vacancy). 
A horizontal dotted line represents the Fermi energy corresponding to ($\Box^{+}$)$_5$(Pb$^{4+}$)$_2$(O$^{2-}$)$_6$+($-e$). 
}
\end{center}
\end{figure}

\begin{figure}[tbp]
\begin{center}
\includegraphics{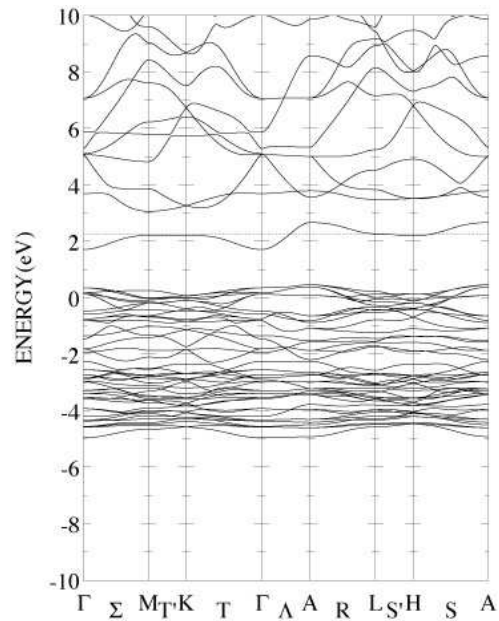}
\caption{\label{fig5} Calculated energy band structure of Ag$_5$$\Box_2$O$_6$ ($\Box$ = Pb vacancy). 
A horizontal dotted line represents the Fermi energy corresponding to (Ag$^{+}$)$_5$($\Box^{4+}$)$_2$(O$^{2-}$)$_6$+($-e$). 
}
\end{center}
\end{figure}

In order to make sure the origin of the conduction band, energy band structures are calculated for $\Box_5$Pb$_2$O$_6$ 
and Ag$_5$$\Box_2$O$_6$ ($\Box$ = vacancy) with the same crystal structure as Ag$_5$Pb$_2$O$_6$ 
and shown in Figs.~\ref{fig4} and \ref{fig5}, respectively. 
SCF calculations are performed for both fictitious systems by keeping charge neutrality. 
Horizontal dotted lines in Figs.~\ref{fig4} and \ref{fig5} represent  
a shifted Fermi energy by assuming ($\Box^{+}$)$_5$(Pb$^{4+}$)$_2$(O$^{2-}$)$_6$+($-e$) and 
(Ag$^{+}$)$_5$($\Box^{4+}$)$_2$(O$^{2-}$)$_6$+($-e$) 
within the rigid-band approximation. 
It is reasonably understood that the conduction band crossing the Fermi energy in Fig.~\ref{fig2} originates in the 
lower branch of the antibonding bands 
of Pb-$6s$ and O-$2p$ in Fig.~\ref{fig4} by judging their dispersion. 
More dispersive nature in Fig.~\ref{fig2} comes from the extra hybridization with the Ag orbitals, which is not involved in $\Box_5$Pb$_2$O$_6$. 
The half-filled antibonding conduction band of Pb-$6s$ and O-$2p$ in Fig.~\ref{fig4} gives just a 2D Fermi surface 
with cylindrical shape and almost no warping along the $c$ direction. 
The warping behavior in Ag$_5$Pb$_2$O$_6$ might come from the transfer integrals between Ag-$4d$ and $5s$ and O-$2p$ orbitals 
as shown in Fig~\ref{fig5}, in which a clear cosine dispersion ($\sim - \cos k_z c$) is seen along $\Gamma$A. 

Calculated total DOS at the Fermi energy is 1.33 states/eV-formula unit. This value corresponds to the electronic specific heat coefficient 
$\gamma=3.13$mJ/K$^2$mol, which gives very small mass enhancement $\lambda=0.09$ by comparing with the experimental 
value $\gamma=3.42$mJ/K$^2$mol.~\cite{Yonezawa2004} 
Calculated Pauli paramagnetic susceptibility is $\chi_0 = 4.30 \times 10^{-5}$emu/mol, which is also comparable 
with the experimental value $(+3.7\pm0.2) \times 10^{-5}$emu/mol. 
It might be hard to evaluate the Stoner enhancement from the comparison 
because the experimental $\chi_0$ may have certain ambiguities due to 
significantly large diamagnetic contribution from the core electrons ($-2.44 \times 10^{-4}$ emu/mol).~\cite{Yonezawa2004}  

\subsection{Fermi surface}

\begin{figure}[tbp]
\begin{center}
\includegraphics{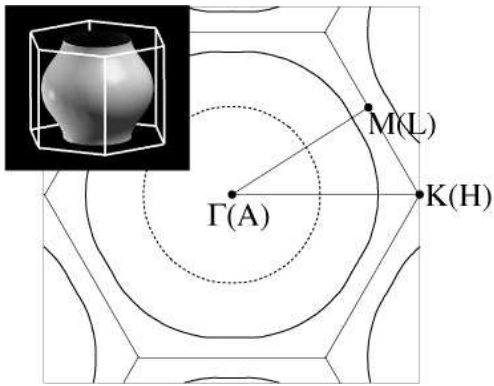}
\caption{\label{fig6} Calculated Fermi surface of the hexagonal Ag$_5$Pb$_2$O$_6$. 
Solid and dotted lines denote the cross sections on the $\Gamma$MK ($k_z=0$) 
and ALH ($k_z=\pi/c$) planes in BZ, respectively. 
Inset is a perspective view of the Fermi surface. 
}
\end{center}
\end{figure}

\begin{figure}[tbp]
\begin{center}
\includegraphics{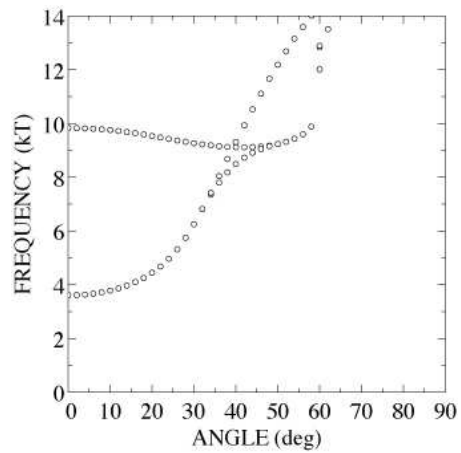}
\caption{\label{fig7} Calculated de Haas-van Alphen frequencies for Fermi surface of the hexagonal Ag$_5$Pb$_2$O$_6$. 
Angle is taken as the polar direction of the magnetic field from $[0001]$ to $[1000]$. 
}
\end{center}
\end{figure}

Obtained band structure for 16$\times$16$\times$16 k-mesh points (417 k points in the irreducible wedge of BZ) 
is fitted with symmetrized star functions by a spline method and used for 
calculations of the Fermi surface and related properties.~\cite{Oguchi1992}  
Figure \ref{fig6} shows Fermi surface of the hexagonal Ag$_5$Pb$_2$O$_6$. 
The cross sections on the $\Gamma$MK ($k_z=0$) and ALH ($k_z=\pi/c$) planes in BZ 
are almost circle, indicating cylindrical shape with large warping along the $c$ direction. 
The Fermi wave numbers from the axis $\Gamma$A along $\Gamma$M, $\Gamma$K, AL and AH are 0.543, 0.551, 0.331 
and 0.332{\AA}$^{-1}$, respectively. The warping along the $c$ direction is of almost perfect cosine shape, 
independent of the lateral components of the wave vector \textbf{k}. 
It is possible to look upon the Fermi surface as a nearly-free-electron-like spherical one deformed by 
overlapping the neighbors at the BZ boundaries of $k_z=\pm\pi/c$. 
Quantum-oscillation measurements such as de Haas-van Alphen (dHvA) and Shubnicov-de Haas effects 
are highly desired since the single crystal is available and can be compared with the present band-theoretical prediction 
to elucidate detailed information of the Fermi surface. 
By applying a magnetic field along the $c$ direction, two quantum-oscillation signals 
should be observed around frequencies of 3.6kT (cyclotron mass $m^* \approx 0.68m$) and 9.8kT ($m^* \approx 1.2m$) 
associated with the extreme Fermi surface cross sections shown in Fig.~\ref{fig6}. 
The large extreme cross section is nearly constant due to its fat belly shape at small polar angles, 
analogous to that expected for a spherical Fermi surface, and
shows discontinuous features because of warping geometry at larger angles than 60$^{\circ}$. 
The small extreme cross section shows strong angle dependence at intermediate polar angles 
and breaks up at 34$^{\circ}$ to two branches, of which one merges into the large one around 50$^{\circ}$. 
The two extreme cross sections corresponding to the dHvA frequencies are depicted in Fig.~\ref{fig7} as a function of the 
polar angle of the applied magnetic field. 

\subsection{Transport properties}

Calculated Fermi velocity is $\langle v_x^2 \rangle^{1/2}=\langle v_y^2 \rangle^{1/2}=3.57\times10^{7}$cm/s 
and $\langle v_z^2 \rangle^{1/2}=2.43\times10^{7}$cm/s. 
According to the Boltzmann theory, the dc conductivity 
is proportional to the Fermi velocity squared with the constant relaxation-time approximation. 
The ratio $\langle v_x^2 \rangle/\langle v_z^2 \rangle \approx 2.16$ may explain small anisotropy of observed resistivity 
$\rho_{c}/\rho_{ab} \approx 2$ at 280K,~\cite{Yonezawa2004} where isotropic diffusive scattering is dominant. 
In Sr$_2$RuO$_4$ and high-$T_c$ superconducting cuprates La$_{1.85}$Sr$_{0.15}$CuO$_4$ and YBa$_2$Cu$_3$O$_7$, 
the anisotropy in the Fermi velocity is quite large, showing strong 2D nature in the electronic structure.~\cite{Oguchi1995}  
It turns out that the hexagonal Ag$_5$Pb$_2$O$_6$ has less 2D character than the ruthenate and cuprates, 
despite of the layered structure with the 1D chains and 2D Kagom\'{e} layers of silver. 

Hall coefficients are also estimated to be $R_{xyz}^H=-12.6\times10^{-10}$m$^3$/C and $R_{yzx}^H=-5.0\times10^{-10}$m$^3$/C. 
Here, the Hall coefficients are defined as 
$R_{\alpha \beta \gamma}^H = E_{\beta}/(j_{\alpha} B_{\gamma})$, 
where $j_{\alpha}$ is a measured electric current in the presence of an electric field $E_{\beta}$ and a magnetic field $B_{\gamma}$. 
The negative sign of the Hall coefficients means that the carrier is an electron. No Hall measurements have been reported so far. 

The temperature dependence of the resistivity may be discussed within the Boltzmann theory 
if the scattering mechanism is dominated by the electron-phonon coupling. 
With use of the coupling taken from the experimental phonon spectra, the resistivity in the high-$T_c$ cuprates 
has been calculated by Allen \textit{et al}.~\cite{Allen1986,Allen1988} and shows $T$-linear dependence, 
which is in good agreement with experiment. 
For Ag$_5$Pb$_2$O$_6$, Yonezawa and Maeno have assumed simple Debye and Einstein model terms in the coupling 
and reproduced 
the $T^2$ dependence of the observed resistivity by adjusting the Debye and Einstein frequencies.~\cite{Yonezawa2004} 
However, the Einstein frequency of higher than 1000K implies the existence of a unique phonon mode 
for understanding the $T^2$ dependence within the electron-phonon mechanism. 

\subsection{Doping effects}

\begin{figure}[tbp]
\begin{center}
\includegraphics{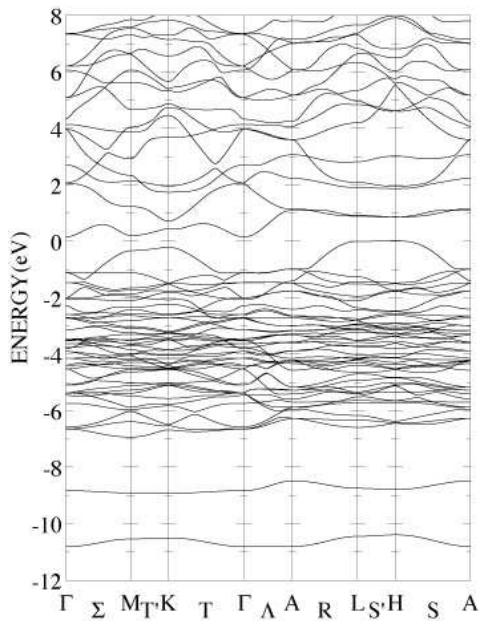}
\caption{\label{fig8} Calculated energy band structure of Ag$_5$PbBiO$_6$. 
The valence band maximum is taken at the origin. 
}
\end{center}
\end{figure}

\begin{figure}[tbp]
\begin{center}
\includegraphics{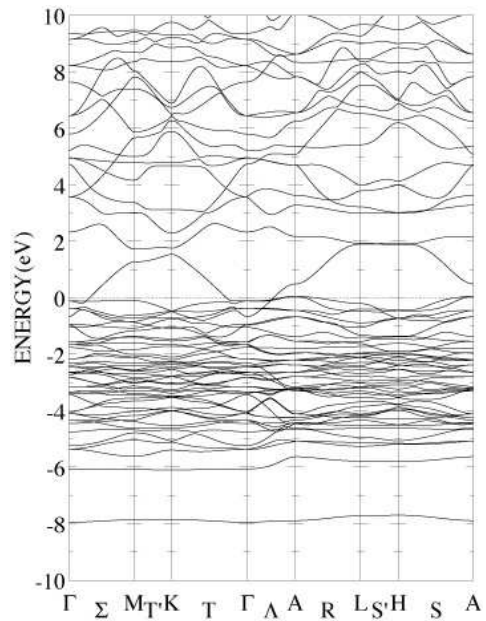}
\caption{\label{fig9} Calculated energy band structure of Ag$_5$PbInO$_6$. 
The Fermi energy is taken at the origin. 
}
\end{center}
\end{figure}

It has been reported that doping of Bi or Cu substituted for the Pb site tends to make the system insulating while 
In doping shows only a moderate increase in the resistivity.~\cite{Bortz1993,Tejada-Rosales2002} 
A naive interpretation for the insulating state can be made for a composition with one valence electron more or less than 
that in Ag$_5$Pb$_2$O$_6$, such as Ag$_5$PbBiO$_6$ and Ag$_5$PbInO$_6$, 
since the conduction band of Ag$_5$Pb$_2$O$_6$ is half filled. 
However, there are no energy gaps below or above the half-filled conduction band as shown in Figs.~\ref{fig2} and \ref{fig3} 
and a simple band-filling picture within a rigid-band model cannot seem to explain the insulating nature. 
Even if there were gaps, it cannot account for the In-doping effect instead. 
A clue to the convincing explanation is the theoretical finding that the conduction band in Ag$_5$Pb$_2$O$_6$ is composed mostly of 
the antibonding state of Pb-$6s$ and O-$2p$, as mentioned above. 
Substitutional doping for the Pb site may give rise to a crucial modification of the antibonding bands. 
We have carried out first-principles electronic structure calculations for Ag$_5$Pb$M$O$_6$ ($M$=Bi, In) 
and results are shown in Figs.~\ref{fig8} and \ref{fig9}. 
In the calculations, crystal structure of Ag$_5$Pb$M$O$_6$ is assumed to be the same as that of Ag$_5$Pb$_2$O$_6$ 
to see doping effects on the electronic structure. 
It is found in Ag$_5$PbBiO$_6$ (Fig.~\ref{fig8}) that there exists one more valence electron occupying the conduction band and 
energy gaps are clearly formed at the middle of the antibonding bands, 
of which the lower occupied and upper empty branches are approximately of the Bi-$6s$ and Pb-$6s$ origin, respectively. 
The gap formation originates in symmetry lowering by substitutional doping for the Pb site. 
On the other hand, no gaps are obtained for Ag$_5$PbInO$_6$ (Fig.~\ref{fig9}) with one less valence electron and 
the conduction and valence bands overlap each other, leading to semi-metallic electronic structure. 
These results of the doping effects are quite consistent with the resistivity data.~\cite{Bortz1993,Tejada-Rosales2002} 
It is, furthermore, necessary to elucidate insulating nature in a Cu-doped system. 

\section{Concluding remarks}

Electronic band structure is obtained for the hexagonal Ag$_5$Pb$_2$O$_6$ by first-principles calculations. 
A half-filled, single conduction band is composed mostly of an antibonding state of Pb-$6s$ and O-$2p$ with 2D nature 
and shows a dispersion along the $c$ direction due to hybridization with the Ag-$4d$ and $5s$ orbitals. 
The resulting Fermi surface has cylindrical shape with large warping along the $c$ direction, 
being regarded possibly as 3D nearly-free-electron-like one. 
Anisotropy in the transport properties is expected to be rather small because of comparable Fermi-velocity components parallel and 
perpendicular to the 2D layer. Observed doping effects on the electronic structure are reasonably understood 
in terms of the filling and gap formation in the conduction bands. 

\begin{acknowledgments}
This work is supported in part by a Grant-in-Aid for Scientific Research in Priority Area (No.16076212) 
and COE Research (No.13CE2002) 
of the Ministry of Education, Culture, Sports, Science and Technology of Japan. 
We thank Yoshiteru Maeno, Singo Yonezawa, and Tatsuya Shishidou for invaluable discussion. 
\end{acknowledgments}

\newpage
\bibliography{Oguchi-APO}

\end{document}